\documentclass[12pt]{article}
\usepackage{epsfig}
\begin{document}
\tolerance=5000
\def\be{\begin{equation}}
\def\ee{\end{equation}}
\def\bea{\begin{eqnarray}}
\def\eea{\end{eqnarray}}
\def\nn{\nonumber \\}
\def\cF{{\cal F}}
\def\det{{\rm det\,}}
\def\Tr{{\rm Tr\,}}
\def\e{{\rm e}}
\def\tr{{\rm tr\,}}

\  \hfill 
\begin{minipage}{3.5cm}
Preprint\\
April 1998 \\
\end{minipage}

\begin{center}

{\large\bf Vacuum  energy for 
 the supersymmetric twisted D-brane in constant electromagnetic field}

\vspace{1.5cm}

{\sc Andrei A. BYTSENKO$^{\ast}$}\footnote{e-mail: 
abyts@fisica.uel.br\,\,\,\,\, On leave from Sankt-Petersburg Technical 
University, Russia},
{\sc Antonio E. GON\c CALVES$^{\ast}$}\footnote{email : 
goncalve@fisica.uel.br} \\ 
\vspace{0.3cm}
{\sl $\ast$ Departamento de Fisica, Universidade Estadual de Londrina, \\
Caixa Postal 6001, Londrina-Parana, BRAZIL}\\

\vspace{0.4cm}

{\sc Shin'ichi NOJIRI$^{\dagger}$}
\footnote{e-mail : nojiri@cc.nda.ac.jp}\\
\vspace{0.3cm}
{\sl $\dagger$ 
Department of Mathematics and Physics National Defence Academy,\\
 Hashirimizu Yokosuka 239, JAPAN}\\

\vspace{0.2cm}

and\\

\vspace{0.2cm}

{\sc Sergei D. ODINTSOV$^{\ddagger}$}
\footnote{e-mail : odintsov@quantum.univalle.edu.co\,\,\,\,\, On leave from
Tomsk Pedagogical University, Russia}\\
\vspace{0.3cm}
{\sl $\ddagger$
Departamento de Fisica, Universidad del Valle,\\
 AA 25360, Cali, COLOMBIA\\}

\vspace{1.5cm}

{\bf ABSTRACT}

\end{center}
We calculate vacuum energy for twisted SUSY D-brane on toroidal background
with constant magnetic or constant electric field. Its behaviour for toroidal 
D-brane (p=2) in constant electric field shows the presence of stable minimum 
for twisted versions of the theory. That indicates such background maybe 
reasonable groundstate.

\noindent
PACS: 04.50.+h, 4.60.-m, 11.25.-w \\

\vspace{1.0cm}

In attempts to understand the 
D-brane theory \cite{2,3,4} study of its vacuum polarization 
may be quite important because of different reasons. In particular, 
calculation of the vacuum energy (or the effective potential) 
(as in the case of p-brane 
\cite{8}) and study their properties may define reasonable background as a 
candidate for groundstate. The investigation of such groundstate can clarify 
if theory contains tachyons in the spectrum or not. 

Toroidal background is one of the simplest candidates for groundstate 
in the theory of extended objects. It is well-known fact that toroidal 
background in the theory of
 $D=11$ supermembrane \cite{DI} is supersymmetric 
one and the vacuum energy is zero on such background. However,
if antiperiodic boundary conditions are chosen for fermions 
(like non-zero temperature) the supersymmetry is usually broken 
and the vacuum energy is not zero. Such twist assignments (antiperiodic 
boundary conditions for all or part of fermions) are 
consistent with classical SUSY transformations so it could be that not 
all supersymmetries of the model are broken (compare with \cite{14}).

Recently the effective potential for SUSY D-brane 
in the constant electromagnetic field has been found \cite{KN} (for the case
of bosonic D-brane see \cite{NO}). 
 However, only periodic boundary conditions 
have been discussed in Ref. \cite{KN}.

The purpose of present letter is to extend results of the paper \cite{KN} to
the case of twisted fermion fields. In particular, we may have such situation 
for the effective potential at non-zero temperature if identify one
of dimensions of torus with temperature.

The supersymmetric D-brane can be given by the following 
Dirac-Born-Infeld-type action (see Refs. \cite{Ab,So,Ag})
\be
\label{DBIS0}
S_D=k\int_0^T d\zeta_0 \int d^p\zeta \e^{-\phi(X)}
\left[\det \left(G_{MN}
\partial_iZ^M \partial_jZ^N +F_{ij}\right)
\right]^{1 \over 2} + S_{WZ}\ ,
\ee
where $\zeta^i$'s are coordinates on the D-brane world 
sheet ($i,j=0,1,\cdots, p$), 
$F_{ij}$ is the electromagnetic field strength on 
the D-brane world sheet:
\be
\label{fstrgth}
F_{ij}=\partial_i A_j - \partial_j A_i \ ,
\ee
and $Z^M$ are the superspace 
coordinates of D-brane
\be
\label{Z}
\{Z^M\}=(X^\mu, \theta^\alpha),
\ee
($\mu, \nu = 0,1,\cdots,9$, $\alpha=1,2,\cdots,32$).
 $\theta$ is a Majorana-Weyl (16 independent 
components) spinor in case of heterotic string, 
a Majorana spinor in case of type IIA superstring 
and are two kinds of Majorana-Weyl (totally 32 components) 
spinors in case of type IIB superstring.
In Eq. (1) $\phi(X)$ is a dilaton field and 
$S_{WZ}$ is Wess-Zumino like term containing 
anti-symmetric tensor fields. Note that this term will not 
be important for our purposes so we dont discuss it below.

We consider the case when the action is given by 
\be
\label{DBIS}
S_D=k\int_0^T d\zeta_0 \int d^p\zeta \e^{-\phi(X)}
\left[\det \left((G_{\mu\nu}+B_{\mu\nu})
\Pi_i^\mu \Pi_j^\nu +F_{ij}\right)
\right]^{1 \over 2}\ ,
\ee
$G_{\mu\nu}$ is the metric of the space-time, 
$B_{\mu\nu}$ is the anti-symmetric tensor 
and 
\be
\Pi_i^\mu \equiv \partial_i X^\mu
- i\bar\theta \Gamma^\mu \partial_i \theta\ ,
\ee
where $\Gamma^\mu=e_a^\mu\gamma^a$ ($e_a^\mu$ is the vielbein 
field and $\gamma^a$'s denote $\gamma$ matrices in 10 
dimensions).  
For simplicity, we assume $G_{\mu\nu}$ and $B_{\mu\nu}$ are 
constants.
The action (\ref{DBIS}) is invariant under the
global super-Poincar\'e transformation
\bea
\label{sP}
\delta X^\mu &=& i\bar\epsilon \Gamma^\mu \theta, \nn
\delta \theta &=& \epsilon.
\eea 
The correspondent transformations for vector and tensor fields 
are not written explicitly.
The system also has 
local reparametrization invariance with respect to 
$\zeta$ and $\kappa$-symmetry.

The reparametrization invariance is fixed by choosing the gauge condition
\be
\label{ggch}
X^i=R_i \zeta^i \ \ (R_0=1),\ \ \ \ i=0,1,\cdots,p\ .
\ee
A fermionic symmetry is fixed by
\be
\label{kappagf}
\Gamma \theta= - \theta\ ,
\ee
where matrix $\Gamma$ squares to 1 and has a vanishing trace; $(1+\Gamma)$ is
a projector which makes a 32-dimensional parameter of $\kappa$-supersymmetry
effectively only 16-dimensional.
Note that in the above gauge, there appear no 
Faddeev-Popov ghosts.

We want to address the stability of  
D-brane when $F_{ij}$ (or 
$B_{ij}$) has nontrivial vacuum expectation value
and study the effective potential, which is defined by
\be
\label{EP}
V=-\lim_{T\rightarrow\infty}{1 \over T}
\ln \int {DX^\bot D\theta^\alpha DA_i \over V_A}\e^{-S_D}\ .
\ee
Here $V_A$ is the gauge volume for the gauge field $A_i$ and 
\be
\label{Xbot}
X^\bot = (X^{p+1},X^{p+2},\cdots,X^9) \ .
\ee
The integration in Eq. (\ref{EP}) with respect to $\theta$ 
should be understand to be integrated over 
the $\theta^\alpha$ space restricted by (\ref{kappagf}).
We now also impose boundary conditions corresponding to the toroidal D-brane.
The twist assignments to $X, \theta, \epsilon$ must be consistent with the
supersymmetry transformations (6). For instance the case when spinor fields are
twisted is admissible as well as the case when the fields are untwisted. Note
that the number of non-isomorphic linear real bundles over the toroidal space
$T^{p+1}$ (the number of spinor fields) is the number of elements in 1st
cohomology group ${\bf H}^1(T^{p+1};{\bf Z}_2)={\bf Z}_2^{p+1}$. Therefore the
number of topologically non-equivalent spinor configurations is $2^{p+1}$ and
we have
\bea
\label{pbc}
&& X^m(\zeta^0, \zeta^1, \zeta^2, \cdots, \zeta^p)
 =X^m(\zeta^0+T, \zeta^1, \zeta^2, \cdots, \zeta^p) \nn
&& =X^m(\zeta^0, \zeta^1+1, \zeta^2, \cdots, \zeta^p) 
 =X^m(\zeta^0, \zeta^1, \zeta^2+1, \cdots, \zeta^p) \nn
&& = \cdots =X^m(\zeta^0, \zeta^1, \zeta^2, \cdots, \zeta^p+1)\,\,\,
(m=p+1,p+2,\cdots,9)\ ,\nn
&& \theta^\alpha(\zeta^0, \zeta^1, \zeta^2, \cdots, \zeta^p)
 =\pm\theta^\alpha(\zeta^0+T, \zeta^1, \zeta^2, \cdots, \zeta^p) \nn
&& =\pm\theta^\alpha(\zeta^0, \zeta^1+1, \zeta^2, \cdots, \zeta^p) 
 =\pm\theta^\alpha(\zeta^0, \zeta^1, \zeta^2+1, \cdots, \zeta^p) \nn
&& = \cdots =\pm\theta^\alpha(\zeta^0, \zeta^1, \zeta^2, \cdots, 
\zeta^p+1)\ ,
\eea
where $\alpha=1,2,\cdots, 16$ for type IIA,\, IIB superstring and 
$\alpha=8$ for heterotic string. Below we only consider the 
Type IIA and IIB superstrings with 
\bea
\label{smpl}
&& \phi=0,\,\,\,\,\, G_{\mu\nu}=\delta_{\mu\nu}, \nn
&& B_{mn}=B_{im}=B_{mi}=0 \  \hskip 0.2cm 
(i=0,1,\cdots,p;\ m,n=p+1, \cdots, 9),
\eea
and assume that anti-symmetric tensor fields in $S_{WZ}$ vanish.
We also divide the anti-symmetric part $\cF_{ij}$ in 
$\hat G_{ij}$ into the sum of 
the classical part $\cF_{ij}^c$ and 
the quantum fluctuation $F_{ij}^q$:
\be
\label{div}
\cF_{ij}=\cF_{ij}^c + F_{ij}^q\ .
\ee
In the following we write $R_i^2\delta_{ij}+\cF_{ij}^c$ 
as $\hat G_{ij}$ and $F_{ij}^q$ as $F_{ij}$.

Quadratic expansion of the action has been done in Ref. \cite{KN}
\bea
\label{quad}
S_2&=&\int_0^T d\zeta_0 \left[
(\det \hat G_{ij})^{1 \over 2}+  
\int d^p\zeta \left\{ {1 \over 2} 
G^{Sij}\partial_i X^\bot
\cdot\partial_j X^\bot-i(\hat G^{-1})^{ij}\bar\theta\Gamma_i 
\partial_j \theta \right. \right. \nn
&& + \left.\left.
\left( -{1 \over 2}(\hat G^{-1})^{jk}(\hat G^{-1})^{li}
+{1 \over 4}(\hat G^{-1})^{ji}(\hat G^{-1})^{lk} 
\right)F_{ij}F_{kl} \right\}\right],
\eea
where
\be
\label{Gtensor}
G^{Sij}\equiv{1 \over 2} 
\left((\hat G^{-1})^{ij}+(\hat G^{-1})^{ji}
\right) \ 
\ee
is defined with the inverse matrix $(\hat G^{-1})^{ij}$ 
of $\hat G_{ij}$. The contribution to the one-loop effective 
potential from 
the $X^\bot$ and the gauge fields  was calculated 
in \cite{NO}. The contribution from fermions 
was found in \cite{KN}: 
\bea
\label{thetaV}
V_\theta &=&-{1 \over 4}
\Tr\ln (\hat G^{-1})^{ij}\Gamma_i\partial_j \nn
&=&-{1 \over 8}\Tr\ln \left((\hat G^{-1})^{ij}\Gamma_i\partial_j 
\right)^2 \nn
&=&-4\Tr\ln g_{ik} (\hat G^{-1})^{ij}
(\hat G^{-1})^{kl}\partial_j 
\partial_l\ .
\eea
Then with the following matrices
\bea
\label{Gtheta}
\hat G_\theta^{ij} &\equiv& g_{kl} G^{Sli} G^{Skj}, \nn
\tilde G_\theta^{\alpha\beta} &=& 
\hat G_\theta^{\alpha\beta} -  
{1 \over \hat G_\theta^{00}} 
\hat G_\theta^{0\alpha} \hat G_\theta^{0\beta},
\eea
 one gets
\be
\label{Vtheta2}
V_\theta=-4\sum_{n_1,n_2,\cdots , n_p=-\infty}^\infty
\left[ 4\pi^2\sum_{\alpha,\beta=1}^p 
\tilde G_\theta^{\alpha\beta} (n_\alpha+\mbox{g}_\alpha)(n_\beta+{\rm g}_
\beta)\right]^{1 \over 2}\ ,
\ee
where ${\rm g}_\alpha,\,{\rm g}_\beta=0$ or $1/2$,\, depending on the
spinor field type choosen in $T^{p+1}$,\,$n_\alpha,\,n_\beta \in {\bf Z}$.

Let us consider explicit examples for 
the choice of electromagnetic background.
First we consider the membrane ($p=2$) with 
the magnetic background
\be
\label{mag}
\cF^c_{0k}=0\ ,\ \ \cF^c_{12}=-\cF^c_{21}=h, 
\ee
and assume
\be
\label{R}
R_1=R_2=R\ .
\ee
Then we obtain
\bea
\label{dete}
\hat G&\equiv& \det \hat G_{ij}= R^4 + h^2, \\
\label{inv}
(\hat G^{-1})^{ij}&=&
\left(
\begin{array}{ccc}
1 & 0 & 0 \\
0 & \hat G^{-1}R^2 & -\hat G^{-1}h \\
0 & \hat G^{-1}h & \hat G^{-1}R^2 
\end{array} \right) \ ,
\eea
and 
\be
\label{Vthetam}
V_\theta = -{16\pi R^3 \over R^4+ h^2}Z\left| \begin{array}{ll}
{\rm g}_1\,\,{\rm g}_2 \\
0\,\,\,0 \\
\end{array} \right|(-1) \ .
\ee
The total one-loop effective potential 
(taking into account the bosonic contribution \cite{NO})
has the following form
$$
V_T=k(R^4+h^2)^{\frac{1}{2}} 
+\frac{8\pi R}{(R^4+h^2)^{\frac{1}{2}}}
\left(Z\left| \begin{array}{ll}
0\,\,0 \\
0\,\,\,0 \\
\end{array} \right|(-1)- Z\left| \begin{array}{ll}
{\rm g}_1\,\,{\rm g}_2 \\
0\,\,\,\,\,\,0 \\
\end{array} \right|(-1)\right)
\mbox{,}
\eqno{(27)}
$$
where the last term gives the contribution from fermions.
Here we have used zeta-function regularization (see ref. \cite{E}
for a review). We define the 2-dimensional Epstein zeta-function associated
with the quadratic form $({\bf n}+{\bf g})^T({\bf n}+{\bf g})=(n_1+{\rm g}_1)
^2+(n_2+{\rm g}_2)^2$ and with $\mbox{Re} s>1$ by the formula
$$
Z\left| \begin{array}{ll}
{\rm g}_1\,\,{\rm g}_2 \\
h_1\,\,h_2\\
\end{array} \right|(s)=\sum_{{\bf n}\in {\bf Z}^2}\mbox{'}
\left[(n_1+{\rm g}_1)^2+(n_2+{\rm g}_2)^2\right]^{-s/2}
$$
$$
\times\exp\left[2\pi i(n_1h_1+n_2h_2)\right]
\mbox{,}
\eqno{(28)}
$$
where ${\rm g}_i$ and $h_i$\,\,$(i=1,2)$ are some real numbers, the prime 
means
omitting the term with $(n_1,n_2)=(-{\rm g}_1,-{\rm g}_2)$ if all the 
${\rm g}_i$ are integers.
For $\mbox{Re} s<1$ the Epstein function is understood to be the analytic 
continuation of the right-hand side of Eq. (28). Defined in such a way, the
Epstein function obeys the explicit form (see Ref. \cite{zuck} for detail)
$$
Z\left| \begin{array}{ll}
{\rm g}_1\,\,{\rm  g}_2 \\
0\,\,\,\,\,\,0\\
\end{array} \right|(-1)=-\frac{1}{2\pi^2}
Z\left| \begin{array}{ll}
0\,\,\,\,\,\,\,\,\,\,\,\,\,\,0 \\
-{\rm  g}_1\,\,-{\rm  g}_2\\
\end{array} \right|(2)
$$
$$
=-\frac{1}{2\pi^2}\times
\left\{\begin{array}{ll}
4\zeta(2)\beta(2), \,\,\,\,\,\,\,\,\, {\rm  g}_1={\rm  g}_2=0, \\
-\eta(2)\beta(2),\,\, \,\,\,\,\, {\rm  g}_1=\frac{1}{2}, {\rm  g}_2=0\,\,
({\rm  g}_1=0, 
{\rm  g}_2=\frac{1}{2})\\
-4\eta(2)\beta(2),\,\,\,\, {\rm  g}_1={\rm  g}_2=\frac{1}{2},
\end{array}\right.
\mbox{,}
\eqno{(29)}
$$
where the function $\zeta(s)$ is the Riemann zeta function, $\eta(s)=
(1-2^{1-s})\zeta(s)$ and $\beta(s)$ is the analytical continuation of the
series sum $\sum_{m=o}^{\infty}(-1)^m(2m+1)^{-s}$ convergent at 
$\mbox{Re}s>0$. Therefore $Z\left|\begin{array}{ll}
0\,\,0 \\
0\,\,0\\
\end{array} \right|(-1)<0$, while $Z\left| \begin{array}{ll}
{\rm  g}_1\,\,{\rm  g}_2 \\
0\,\,\,\,\,\,0\\
\end{array} \right|(-1)>0$ for ${\rm  g}_1=\frac{1}{2}\,\, ({\rm  g}_2=
\frac{1}{2})$ or for ${\rm g}_1={\rm  g}_2=\frac{1}{2}$.

Rescaling
$$
R\rightarrow h^{1 \over 2}R\ ,\ \ \ 
V_T\rightarrow h^{-{1 \over 2}}V_T\ ,
\eqno{(30)}
$$
we obtain
$$
V_T=h^{\frac{3}{2}}(R^4+1)^{\frac{1}{2}}+
\frac{8\pi R}{(R^4+1)^{\frac{1}{2}}}
\left(Z\left| \begin{array}{ll}
0\,\,0 \\
0\,\,\,0 \\
\end{array} \right|(-1)
-Z\left| \begin{array}{ll}
{\rm g}_1\,\,{\rm g}_2\\
0\,\,\,\,\,\,0 \\
\end{array} \right|(-1)\right)
\mbox{.}
\eqno{(31)}
$$
Therefore if ${\rm g}_1={\rm g}_2=0$ then $V_T=h^{3/2}(R^4+1)^{1/2}$, but
if $({\rm g}_1,{\rm g}_2)\neq (0,0)$ one gets

$$
V_T=h^{\frac{3}{2}}(R^4+1)^{\frac{1}{2}}-\frac{4R}{\pi(R^4+1)^{\frac{1}{2}}}
\times\left\{\begin{array}{ll}
4\zeta(2)\beta(2)+\eta(2)\beta(2),\,\, \,\,\,\,\, {\rm  g}_1=\frac{1}{2}, 
{\rm  g}_2=0 \\
\:\:\:\:\:\:\:\:\:\:\:\:\:\:\:\:\:\:\:\:\:\:\:\:\:\:\:\:\:\:\:\:\:\:\:\:\:
\:\:\:\:\:\:\:\:\:({\rm  g}_1=0, {\rm  g}_2=\frac{1}{2}),\\
4\zeta(2)\beta(2)+4\eta(2)\beta(2),\,\,\,\, {\rm  g}_1={\rm  g}_2=\frac{1}{2}.
\end{array}\right.
\eqno{(32)}
$$

For $R\rightarrow 0$ we have $V_T=h^{3/2}+{\cal O}(R)$, while for 
$R\rightarrow\infty$, \,$V_T=h^{3/2}R^2+{\cal O}(R^{-1})$. It can be shown that
the one-loop effective potential has stable minimums iff 
$({\rm g}_1,{\rm g}_2)\neq(0,0)$.

Let us find now the one-loop effective 
potential of the membrane 
in the constant electric background where
$$
\cF_{0k}=e\ ,\ \ \cF_{kl}=0\ ,\ \ R_1=R_2=R\ .
\eqno{(33)}
$$
Then we find
$$
\hat G \equiv \det \hat G_{ij}=R^4 + 2e^2 R^2,
\eqno{(34)}
$$
$$
(\hat G^{-1})^{ij}= \hat G^{-1}
\left(
\begin{array}{ccc}
R^4 & -eR^2 & -eR^2 \\
eR^2 & R^2+e^2 & -e^2 \\
eR^2 & -e^2 & R^2+e^2 
\end{array}
\right) \ .
\eqno{(35)}
$$
Now the total one-loop effective 
potential (taking into account the 
bosonic contribution found in Ref. \cite{NO}) is given by
$$
V_T = k(R^4+2e^2)^{1 \over 2} 
+ {8\pi(R^2 + e^2)^{1 \over 2} 
\over R^2}\left[ 
{\cal Z}_0\left(-1;-{e^2 \over R^2 + e^2}\right)
-{\cal Z}_{\bf g}\left(-1;-{e^2 \over R^2 + e^2}\right)\right]\ .
\eqno{(36)}
$$
The function ${\cal Z}_{\bf g}(s;a)$ is defined by the relation for 
$\mbox{Re} s>1$
$$
{\cal Z}_{\bf g}(s;a)=\sum_{{\bf n}\in{\bf Z}^2}\left[
({\bf n}+{\bf g})^T{\cal A}({\bf n}+{\bf g})\right]^{-s/2}
\mbox{,}
\eqno{(37)}
$$
where ${\cal A}=\left( \begin{array}{ll}
1 & a \\ a & 1 \\ \end{array} \right)$,
$a$ is a real number, i.e. $a^2>0$. Use the Mellin representation
$$
{\cal Z}_{\bf g}(s;a)=\frac{1}{\Gamma(\frac{s}{2})}\sum_{{\bf n}\in{\bf Z}^2}
\int_0^{\infty}dt\cdot t^{s/2-1}\exp\left[-t({\bf n}+{\bf g})^T{\cal A}
({\bf n}+{\bf g})\right]
\mbox{,}
\eqno{(38)}
$$
and the Poisson resummation formula yields the following expression
$$
{\cal Z}_{\bf g}(s;a)=\sqrt{\pi}(1-a^2)^{\frac{1-s}{2}}\frac{\Gamma(\frac{s-1}
{2})}{\Gamma(\frac{s}{2})}
Z\left| \begin{array}{ll}
{\rm g} \\ 0 \\ \end{array} \right|(s-1)
+\frac{2\pi^{\frac{s}{2}}}{\Gamma(\frac{s}{2})}\sum_{n_1=1}^{\infty}
\sum_{n_2=-\infty}^{\infty}\cos\left[2\pi n_1{\cal G}(an_2)\right]
$$
$$
\times\left[\frac{(n_2+{\rm g}_2)\sqrt{1-a^2}}{n_1}\right]^{\frac{1-s}{2}}
K_{\frac{1-s}{2}}\left(2\pi n_1(n_2+g_2)\sqrt{1-a^2}\right)
\mbox{.}
\eqno{(39)}
$$
In Eq. (39) $K_{\nu}(z)$ are the modified Bessel functions and ${\cal G}
(an_2)=a(n_2+{\rm g}_2)+{\rm g}_1$. In the case ${\bf g}=(0,0)$ the ${\cal Z}_0(s;a)$ is
given by the Chowla-Selberg formula \cite{chow49-35-317}
$$
{\cal Z}_{\bf 0}(s;a)=2\zeta(s)+\frac{2\sqrt{\pi}(1-a^2)^{\frac{1-s}{2}}}
{\Gamma(\frac{s}{2})}\zeta(s-1)\Gamma(\frac{s-1}{2})
$$
$$
+\frac{8\pi^{\frac{s}{2}}(1-a^2)^{\frac{1-s}{4}}}{\Gamma(\frac{s}{2})}
\sum_{n=0}^{\infty}n^{\frac{s-1}{2}}\sigma_{1-s}(n)\cos(n\pi a)
K_{\frac{s-1}{2}}\left(2n\pi\sqrt{1-a^2}\right)
\mbox{,}
\eqno{(40)}
$$
where $\sigma_s(n)\equiv\sum_{d|n}d^s$, sum over the $s$-powers of the divisors
of $n$. Finally for $s=-1$ we have
$$
{\cal Z}_{\bf g}(-1;a)=-\frac{1-a^2}{4\pi^2}Z\left| \begin{array}{ll}
\,\,\,\,0 \\ -{\rm g} \\ \end{array} \right|(3)
-\frac{1}{\pi}\sum_{n_1=1}^{\infty}\sum_{n_2=-\infty}
^{\infty}\cos\left[2\pi n_1{\cal G}(an_2)\right]
$$
$$
\times\left[\frac{(n_2+{\rm g}_2)\sqrt{1-a^2}}{n_1}\right]
K_{1}\left(2\pi n_1(n_2+{\rm g}_2)\sqrt{1-a^2}\right)
\mbox{,}
\eqno{(41)}
$$
$$
{\cal Z}_{\bf 0}(-1;a)=-\frac{1}{6}-\frac{1-a^2}{4\pi^2}\zeta(3)
$$
$$
-\frac{4}{\pi}\sqrt{1-a^2}
\sum_{n=0}^{\infty}\frac{\sigma_2(n)}{n}\cos(n\pi a)
K_1\left(2n\pi\sqrt{1-a^2}\right)
\mbox{.}
\eqno{(42)}
$$
In Eq. (41) the functional equation (analytic continuation)
$$
\pi^{-\frac{s}{2}}\Gamma\left(\frac{s}{2}\right)Z\left| \begin{array}{ll}
{\rm g} \\ h \\ \end{array} \right|(s)=\pi^{\frac{s-1}{2}}\Gamma\left(
\frac{1-s}{2}\right)
\exp(-2\pi i {\rm g}h)Z\left| \begin{array}{ll}
h \\ -{\rm g} \\ \end{array} \right|(1-s)
\mbox{,}
\eqno{(43)}
$$
has been used. A simple calculation along the above line gives 
$$
Z\left| \begin{array}{ll}
\,\,\,\,0 \\ -{\rm g} \\ \end{array} \right|(3)=
\left\{\begin{array}{ll}
2\zeta(3), \,\,\,\,\,\,\,\,\, {\rm g}_2=0, \\ 
-2\eta(3),\,\,\,\,\, {\rm g}_2=\frac{1}{2}, \\ 
\end{array} \right.
\mbox{.}
\eqno{(44)}
$$
and $Z\left| \begin{array}{ll}
0 \\ 0 \\ \end{array} \right|(3)>0$, while $Z\left| \begin{array}{ll}
\,\,\,\,\,\,0 \\-1/2 \\ \end{array} \right|(3)<0$.

When $R \rightarrow 0,\,\,(a\rightarrow -1)$ and 
$({\rm g}_1,{\rm g}_2)\neq (0,0)$ the one-loop effective potential behaves as
$$
V_T=\frac{8\pi e}{R^2}\left[{\cal Z}_0(-1;-1) - {\cal Z}_{\bf g}(-1;-1)
\right]+{\cal O}(1)
\mbox{,}
\eqno{(45)}
$$ 
where ${\cal Z}_{\bf g}(-1;a\rightarrow -1)<0$ for ${\rm g}_1={\rm g}_2$ and 
${\cal Z}_{\bf g}(-1;a\rightarrow -1)>0$ for ${\rm g}_1-{\rm g}_2=\pm 1/2$.
On the other hand, when $R\rightarrow \infty\,\,(a\rightarrow 0)$ we have
$$
V_T= kR^2+{\cal O}(R^{-1})
\mbox{.}
\eqno{(46)}
$$
Thus using Eqs. (45) and (46) one can show that there are minimums of
the function $V_T$. This should be compared with the 
effective potential of the bosonic D-membrane in the electric 
background, which is unstable. Thus supersymmetric D-brane in constant 
electric field becomes the stable one unlike the case of the bosonic 
D-brane. That indicates such background maybe reasonable candidate 
for groundstate. Then detailed quantization of D-brane on such 
background deserves further study.

\vspace{0.5cm}
{\bf ACKNOWELEDGMENTS}
\vspace{0.5cm} 

We would like to thank A.Sugamoto for helpful discussions.
The research of A.A. Bytsenko was supported in part by CNPq grant (Brazil), 
by Russian Foundation for Basic research (grant No. 98-02-18380-a) and by 
GRACENAS (grant No. 6-18-1997). Research of S.D. Odintsov has been supported 
in part by COLCIENCIAS (Colombia).

\end{document}